\documentclass[12pt]{article}

\usepackage{graphicx}
\usepackage{amsmath}
\usepackage{amsfonts}

\usepackage{textcomp}
\usepackage{enumerate}   

\usepackage{enumitem}
\usepackage[round]{natbib}
\usepackage{authblk}

\begin{document}

\title{Dynamical effects of inflation in ensemble-based data assimilation under the presence of model error}
\author[1,2,*]{Scheffler Guillermo}

\author[3,4]{Carrassi Alberto}
\author[1,2]{Ruiz Juan}
\author[5,6,2]{Pulido Manuel}

\affil[1]{\footnotesize Centro de Investigaciones del Mar y la Atm\'osfera, CONICET-UBA, Departamento de Ciencias de la Atm\'osfera y los Oc\'eanos, FCEyN, UBA, Buenos Aires, Argentina}
\affil[2]{UMI-IFAECI 3355 CNRS-CONICET-UBA, Buenos Aires, Argentina}
\affil[3]{Dept. of Meteorology and National Centre for Earth Observations, University of Reading, UK}
\affil[4]{Mathematical Institute, Utrecht University, Utrecht, the Netherlands}
\affil[5]{Departamento de F\'{\i}sica, FaCENA, UNNE, Corrientes, Argentina}
\affil[6]{CONICET, Corrientes, Argentina}
\affil[*]{Corresponding author: Guillermo Scheffler, guillermo.scheffler@cima.fcen.uba.ar}

\date{October 2021}

\maketitle
\begin{abstract}
The role of multiplicative and additive covariance inflation on ensemble dynamics under the presence of model errors is examined. We show that multiplicative inflation significantly impacts the alignment of ensemble anomalies onto the backward Lyapunov vectors. Whereas the ensemble is expected to collapse onto the subspace corresponding to the unstable portions of the Lyapunov spectrum, the use of multiplicative inflation contributes to retain anomalies beyond that span. Given that model error implies that analysis error is not fully confined on the local unstable subspace, this uncovered feature  of  multiplicative inflation is of paramount importance for an optimal filtering. We propose hybrid schemes whereby additive perturbations complement multiplicative inflation by suitably increasing the dimension of the subspace spanned by the ensemble. The use of hybrid schemes improves analysis RMSE on the Lorenz-96 model compared to the use of multiplicative inflation alone, emphasizing the role of model dynamics when designing additive inflation schemes.
\end{abstract}

\section{Introduction}
{\label{sec_intro}}
  Dynamical properties of forecast models play a key role on the performance of data assimilation (DA). For deterministic dynamics in a perfect-model scenario with linear dynamical and observational models, the span of the forecast error covariances of the Kalman filter (KF) and smoother (KS) aligns asymptotically  with the subspace spanned by the backward Lyapunov vectors (BLVs) associated to non-negative Lyapunov exponents \citep{gurumoorthy17,bocquet2017a,bocquet17b}. Under the same deterministic and perfect model scenario, the connections between the filter/smoother error covariance and the underlying model Lyapunov spectrum have also been numerically established for the deterministic versions of the ensemble Kalman filter (EnKF) and smoother (EnKS), whenever the error regime is bound to be weakly nonlinear \citep{bocquet17b}. As a direct consequence, the minimal effective number of ensemble members required in an EnKF to achieve optimal performance is equal to the number of non-negative Lyapunov exponents (LEs). These findings extend to dynamical systems displaying a degeneracy in the spectrum of LEs, but with some caveats. In some coupled ocean-atmosphere models, the coupling manifests itself by the appearance of a quasi-flat portion of the LE spectrum with many LEs very close (within machine accuracy) to zero \citep{tondeur2020temporal}. In this case, the error covariance in a strongly coupled EnKF asymptotically aligns with the subspace spanned by the BLVs associated to all unstable plus all of the quasi-neutral modes, and the effective number of ensemble members must include all of those additional (possibly asymptotically stable) quasi-neutral LEs \citep{carrassi2020data}.
The DA framework known as assimilation in the unstable subspace (AUS) exploits these features explicitly for the design of a data assimilation methodology. In AUS the analysis solution is confined to the subspace spanned by the unstable and neutral BLVs (see e.g. \citealp{palatella13} and references therein).

In the presence of random model errors the role of model dynamics on filter behavior and performance is less evident. In unstable stochastic linear systems with Gaussian errors, \citet{grudzien18a} have proved that, because of the injection of random model error on all of the model degrees of freedom, the error projection onto stable BLVs is not longer dampened asymptotically to zero, even tough it remains bounded. Nevertheless, the size of those bounds, which depends on the amplitude and rank of the model error covariance matrix but also on the variability of the local LEs (LLEs), can be impractically too large. The mechanism is driven by events in which an asymptotically stable mode (with ${\rm LLE}<0$) experiences instantaneous error growth, thus the amplitude and frequency of these local bursts of instabilities impact on the bounds. Naturally, the weakly stable modes (those having an asymptotic ${\rm LE}\approx0$) are the most relevant as their chances to be instantaneously unstable are larger. Unlike the deterministic system case, this behaviour implies the need to increase the effective number of ensemble members in an EnKF to include as many additional members as the number of weakly stable modes of high variability \citep{grudzien18b}. 

Nevertheless, the increase of ensemble members only amounts to a necessary condition and it does not guarantee filter convergence. Local bursts of instability on the weakly stable modes contribute to enhance the so called  {\it upwelling} phenomenon \citep{grudzien18b}. This is, errors in the unfiltered subspace is upwelled to the filtered subspace in a reduced-rank filter, even if the unfiltered subspace only contains asymptotically stable directions. The effect is inherent to the reduced-rank formulation, thus is present also in deterministic systems, but it is much more pronounced in the presence of model errors. The upwelling mechanism provides a rationale and justifies the use of {\it covariance inflation} \citep[see, {\it e.g.}][]{whitaker12} to counteract the filter underestimation of the actual forecast covariance caused by both sampling and model errors.

Covariance inflation is commonly applied under two forms: {\it multiplicative} and {\it additive}. {\it Multiplicative inflation} involves re-scaling the ensemble anomalies ({\it i.e.}, their deviations from the mean) by a scalar factor \citep{anderson99}. Alternatively, {\it additive inflation} consists in adding random noise to each ensemble member \citep{hamill05}. \citet{whitaker12} suggest that multiplicative inflation methods may be more suitable for the treatment of sampling errors associated to the observation network density, whereas additive inflation can alleviate systematic model errors.  
In both cases, one has to tune the key parameters ({\it e.g.}, the amplitude of the rescaling factor in the multiplicative case and the amplitude of the noise in the additive case). This is a very costly procedure in high-dimensional spaces and not exempt of inadequacies.    
Adaptive inflation schemes attempt to estimate those parameters online as part of the DA process and usually rely on observation-based diagnoses \citep{li09,miyoshi11} or on estimating maximum likelihood parameters for covariance scaling \citep{anderson09,zheng2009,el18}. Different approaches to adaptively inflate covariance matrices are compared by \citet{el19} showing that the combination of multiple adaptive inflation schemes can target different sources of errors such as sampling errors and model deficiencies. A comprehensive review of adaptive covariance inflation schemes can be found in \citet{raanes19}. 

Nonetheless, covariance inflation schemes do not take into account model dynamics explicitly in their formulation, nor it is fully understood how the aforementioned connection between the underling model instabilities and the filter error description behaves in the presence of inflation schemes. This is particularly noteworthy for multiplicative inflation schemes whereby the span of the forecast error covariance is in principle not affected by the inflation. In contrast, additive inflation schemes introduce stochastic noise that effectively perturb the dynamical evolution of the ensemble anomalies.

In this context, perturbations from additive inflation that lay within the stable subspace are expected to vanish asymptotically similarly to what happens to model noise. However, transient instabilities may drive additive perturbations onto the filtered (presumably unstable-neutral) subspace. This mechanism puts in evidence how the efficacy of additive inflation is affected by the underlying model instability properties. We shall show however that such an interplay exists, contrary to our original belief, also for multiplicative inflation. 

Our work aims at studying in depth the relation between nonlinear model dynamics and the functioning of the ensemble Kalman  filter upon the presence of model error and in particular the interplay between model instabilities and the inflation scheme. We demonstrate that, by the sole introduction of multiplicative covariance inflation, the EnKF may span a subspace beyond the most unstable directions. New inflation methods are proposed that focus on decomposing the ensemble anomalies to properly account for model errors upwelled from asymptotically stable modes. This is achieved by targeting inflation of ensemble anomalies based on their degree of alignment onto unstable or weakly stable BLVs respectively. 

\section{Problem and experimental settings}
\label{sec:ProbSet}
    Let the state and observations be defined through the non-linear state space model:
    \begin{equation}
        \mathbf{x}_{k} = \mathcal{M}_k(\mathbf{x}_{k-1})+ \boldsymbol{\eta}_{k}, \hspace{0.75cm} k=1,2,...
        \label{eq_stat}
    \end{equation}
    \begin{equation}
        \mathbf{y}_{k} = \mathbf H\mathbf{x}_{k}+ \boldsymbol{\nu}_{k}. \hspace{1.7cm} k=1,2,...
        \label{eq_obs}
    \end{equation}

    The  state $\mathbf{x}_k \in \mathbf{R}^n$ evolves according to the non-linear autonomous dynamical model $\mathcal{M}$ from time $t_{k-1}$ to $t_{k}$. Observations $\mathbf{y}_k \in \mathbf{R}^p$ are related to the state through an observation operator matrix, $\mathbf H\in \mathbf{R}^{ p \times n}$, assumed here, for the sake of simplicity, to be linear. The model error, $\boldsymbol{\eta}_k$, and the observational error $\boldsymbol{\nu}_k$ are assumed to be mutually independent and to be Gaussian distributed with zero-mean  and covariance $\mathbf{Q}$ and $\mathbf{R}$ respectively. These covariance matrices are assumed to be known. State estimation is performed using the ensemble transform Kalman filter \citep[ETKF,][]{hunt07} in which the prediction covariance is approximated using an ensemble representation. The ensemble members are evolved using the non-linear dynamical model to obtain a forecast ensemble ${\{\mathbf{x}^{\rm{f}(i)}_k |~ i=1,2,\ldots,N \}}$:

    \begin{equation}
        \mathbf{x}_{k}^{\rm{f}(i)} = \mathcal{M}_k(\mathbf{x}_{k-1}^{\rm{a}(i)}), \hspace{1.5cm} k=1,2,... , \hspace{0.5cm} i=1,2,...,N
        \label{eq_for}
    \end{equation}
where $\mathbf{x}_{k}^{\rm{a}(i)}$ is the $i-th$ member analysis state at the $k-th$ assimilation cycle.

We shall consider a trajectory generated by Eq.~\eqref{eq_stat} as the ``true state''. Synthetic observations are generated with Eq.~\ref{eq_obs} and the hidden true state is then estimated from the set of noisy observations using the ETKF. The ensemble members are propagated in time using the deterministic model, Eq.~\eqref{eq_for}, in which stochastic model error is not included. 

Experiments are performed using the Lorenz-96 system \citep{lorenz98} with $n=40$ variables and forcing $F=8$. In this configuration the dynamical system contains $n_0=14$ non-negative LEs. The model is numerically integrated with a timestep, $\delta t=0.05$, for $1,250$ model time units ({\it i.e.} for $25,000$ time steps). The state is completely observed, {\it i.e.} $p=n$, at every model time step, with an observational error covariance set to ${\mathbf{R} = 0.05^2 \mathbf{I}_n}$. The choice of such an observational constraint is intentional, and it is required to maintain the evolution of the state estimation error within a linear or weakly non-linear regime, such that the dynamics of the system could be reasonably modelled using the BLVs.

The model error covariance is a scaled version of the covariance used in \citet{grudzien18b},
\begin{equation}
\label{eq:Q}
\mathbf{Q} \overset{\Delta}{=}\sigma_q^2
\begin{pmatrix}
0.5 & 0.25 & 0.125 & 0 & \cdots & 0.125 & 0.25 \\
0.25 & 0.5 & 0.25 & 0.125 & \cdots & 0 & 0.125 \\
\vdots & 0.25 & 0.5 & 0.25  & \ddots & \ddots & \vdots \\
0.125 &  \ddots  & \ddots & \ddots  & \ddots & 0.5 & 0.25 \\
0.25 & 0.125 & 0 &\cdots & \cdots & 0.25 & 0.5
\end{pmatrix}
.
\end{equation}

The experiments were conducted with an ensemble size of $N=32$ members, with initial state drawn from the normal distribution, ${\mathcal{N}(\mathbf{x}_{0},\sigma_b^2\mathbf{I}_n)}$. After an initial exploratory analysis, we set $\sigma_b=0.5$ but results were insensitive to the choice of $\sigma_b$ after a spin-up period. The adaptive multiplicative inflation scheme by \citet{miyoshi11} is used. In all of the experiments, the first $2,500$ assimilation cycles are discarded from the diagnostics calculations. For the sake of robustness,  all results are averaged over $10$ repetitions of the assimilation experiments, each one using different initial ensemble and different observational error samples.

    \section{Ensemble and unstable subspaces with model error and the role of multiplicative inflation}
    {\label{sec_baseline}}

\begin{figure}[!hbt]
	\begin{center}
    \includegraphics[width=\columnwidth]{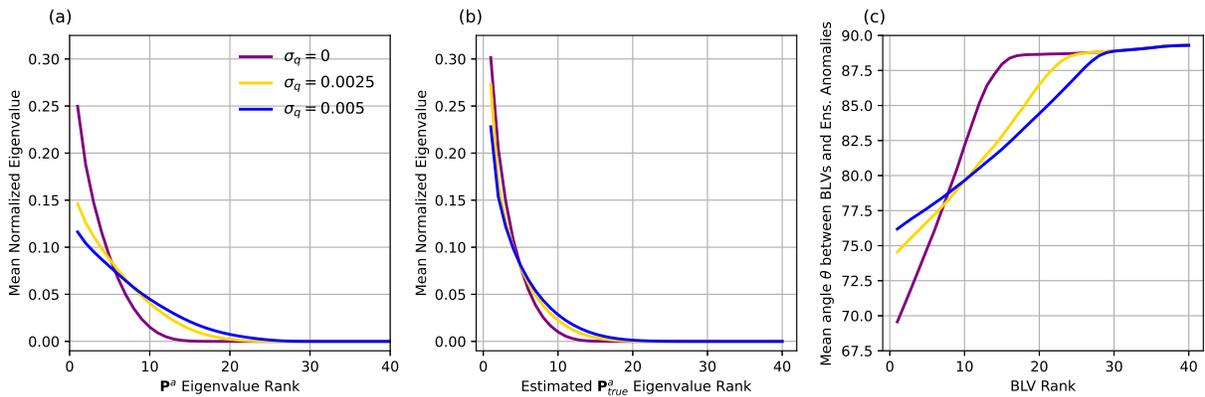}
	\caption{(a): Normalized mean eigenvalues of the ETKF analysis error covariance matrix $\mathbf{P}^{\rm a}$ for different model error amplitudes  $\sigma_q=(0;0.0025;0.005)$. Background covariances were artificially inflated using a multiplicative adaptive covariance scheme \citep{miyoshi11}. (b): Normalized  mean eigenvalues of the covariance of analysis errors $\mathbf{\epsilon}^{\rm a}_k=\mathbf{x}_k-\mathbf{\overline{x}}_k$ calculated over 50 repetitions with same assimilation set up as (a). (c): Time and ensemble averaged angle $\theta$ of ensemble anomalies onto the backward Lyapunov vectors (BLVs).  Multiplicative adaptive inflation coefficients averaged over time were $(1.01;1.114; 1.182)$ respectively.}
	\label{pevq}
	\end{center}
	\end{figure}

    Experiments using a perfect model scenario ({\it i.e.} $\sigma_q=0$) confirm the asymptotic alignment of the ensemble subspace with the unstable-neutral subspace. As shown in Fig.~\ref{pevq}a (purple line),  eigenvalues of the analysis error covariance matrix $\mathbf{P}^{\rm a}$ vanish beyond the rank $r>n_0=14$. This result is in agreement with that in \citet{bocquet17b}  even though a different deterministic square-root EnKF \citep[see {\it e.g.}][]{tippett03,bocquet15} is used in that work. The incorporation of additive model error results in a flattening of the eigenspectrum of $\mathbf{P}^{\rm a}$  and it is no longer bounded to the first $14$ eigenvalues ({\it cf} blue and yellow lines in Fig.~\ref{pevq}a). It is worth mentioning that the effect of model error on the rank of the covariance matrix is only present in the ensemble-based covariance, and not in the (generally unknown) actual error covariance matrix. To see this we show the spectrum of eigenvalues for both matrices in Fig.1a and 1b. The actual covariance matrix of analysis errors $(\mathbf{P}^{\rm a}_{\rm true})$ is  empirically estimated from  analysis errors $\boldsymbol{\varepsilon}^{a}_k=\mathbf{\overline{x}}_k^{a}-\mathbf{x}_k$ over experiment repetitions and averaged over time, where $\mathbf{\overline{x}}_k^{a}$ represents the analyzed state at time $k$ . To ensure a proper rank estimation, assimilation experiments were repeated 50 times for this particular calculation, instead of 10, using the same repetition methodology described on the previous section. Covariance $\mathbf{P}^{\rm a}_{true}$ also exhibits a flattening of the eigenspectrum as model error increases ({\it cf} blue and yellow lines in Fig.~\ref{pevq}b), however the presence of model error does not increase the rank of $(\mathbf{P}^{\rm a}_{\rm true})$ as drastically as in the ensemble-based $\mathbf{P}^{\rm a}$. This suggests that the dynamics of actual analysis errors are not as sensitive to model errors configurations as ensemble based covariance $\mathbf{P}^{\rm a}$.  From now on, we focus on ensemble based covariances $\mathbf{P}^{\rm a}$ to understand the role of model errors and multiplicative inflation on ensemble dynamics.

\subsection{The role of model error}


 To properly examine the distinction between the perfect and imperfect model cases, we analyze the alignment of ensemble anomalies $\mathbf{A}^{(i)}_k=\mathbf{x}_k^{\rm{f(i)}}-\mathbf{\overline{x}}^{\rm f}_k$ with the BLVs computed through successive orthonormalization \citep{legras96}.
Following \citet{bocquet17b} we define the mean angle between ensemble anomalies $\mathbf{A}\in\mathbb{R}^{n\times N}$ such that $\mathbf{P}^f=(N-1)^{-1}\mathbf{A}\mathbf{A}^\top$ and the $j-th$ BLV, stored as the $j-th$ column of the matrix $\mathbf{E}^j_k$ at time $t_k$ as 

    \begin{equation}
        \theta^j_k=\frac{1}{N}\sum^N_{i=1}\theta^{i,j}_k = \frac{1}{N}\sum^N_{i=1} \arccos \Bigg\{ \frac{(\mathbf{E}^j_k)^\top\mathbf{A}^{(i)}_k}{\lVert\mathbf{A}^{(i)}_k\rVert}\Bigg\} , 
    \end{equation}
    with $\theta^j_k \in [0,\pi]$  and  $1\leq j \leq N$, and where $\mathbf{A}^{(i)}_k$ is the $i-th$ column of the anomaly matrix $\mathbf{A}$ such that $\mathbf{P}^f=\mathbf{A}\mathbf{A}^\top$

The time averaged angle $\theta$ is shown in Fig.~\ref{pevq}c. In the perfect model case (purple line), we see a clear cut-off in the projections around rank $n_0=14$, marking the net difference between the stable and the unstable-neutral portions of the Lyapunov spectrum. The ensemble anomalies have negligible projections onto the stable subspace. 
Cut-offs are visible in the experiments with model error too, but they occur now in correspondence to larger BLV ranks directly proportional to the size of the model error ({\it i.e.}, the larger the model error the higher the rank).
The projection of the Kalman filter $\mathbf{P}^{\rm a}$ onto the stable subspace does no longer vanish. The filtered error is present in a few weakly-stable modes. Its number grows with the model error amplitude. This behaviour is consistent with what is anticipated in \citet{grudzien18a} for linear systems. Figure~\ref{pevq}c reveals also that the emergence of anomalies in weakly-stable directions is counter-balanced by a reduction in the projection onto leading BLVs ($r \leq 7$). This effect is reminiscent of the role of growing non-linearities in a perfect model setting as described in \citet{bocquet17b}.

\subsection{The role of inflation}

We investigate now the impact of covariance inflation on the span and rank of the EnKF anomalies matrix, and on how these anomalies project onto the BLVs. 
The time-average multiplicative covariance inflation factors for the three experiments are reported in Table~\ref{tab:base_exp}, together with the corresponding RMSE of the analysis. 
The inflation factor in the adaptative scheme by \citet{miyoshi11} is calculated based on a Bayesian formulation. It is therefore minimal in the perfect model case, and it increases proportionally to the degree of the model error.

\begin{table}[h!]
\centering
\normalsize
\caption{Analysis RMSE and time-average multiplicative adaptive inflation coefficient, $\alpha$, for different levels of model error.}
\label{tab:base_exp}
\begin{tabular}{|l|c|c|c|}
			\hline
			Model Error Amplitude & $\sigma_q=0$ & $\sigma_q=0.0025$ & $\sigma_q=0.005$ \\
			\hline
			Analysis RMSE & 0.008 & 0.014 & 0.016\\
			\hline
			Time-mean inflation coefficient - $\alpha$ & 1.010 & 1.114 & 1.182\\
			\hline

\end{tabular}
\end{table}

\begin{figure}[!hbt]
	\begin{center}
    \includegraphics[width=\columnwidth]{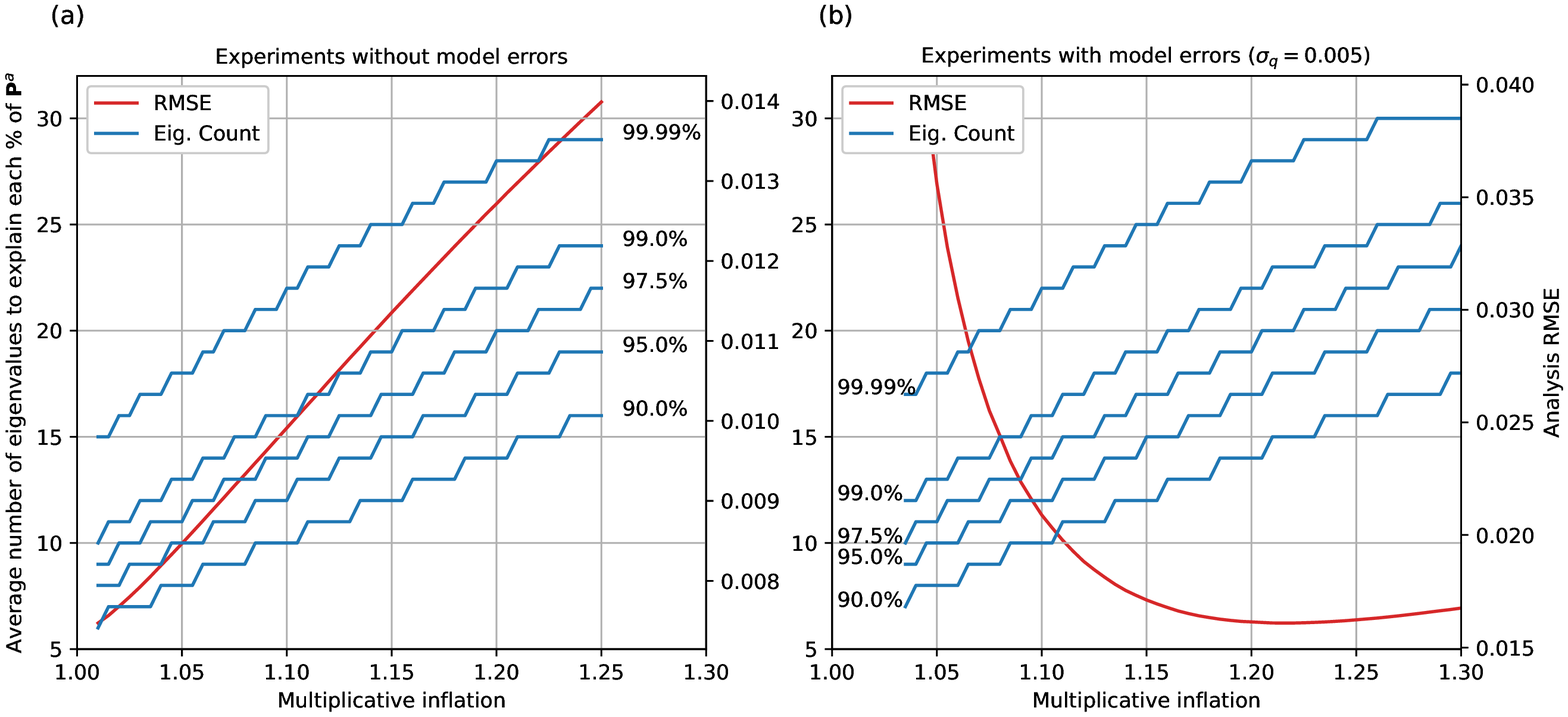}
	\caption{Average number of eigenvalues of $\mathbf{P}^{\rm a}$
required to explain a portion of the full analysis error variance (blue) and the analysis RMSE (red) as a function of the fixed multiplicative covariance inflation. Results are shown for the perfect model case (a), model error case with $\sigma_b=0.005$ (b). Results are averaged over $15,000$ assimilation cycles. Multiplicative inflation values $\alpha$ that are not large enough to avoid filter divergence were intentionally excluded.}
	
	\label{infl_rmse}
	\end{center}
	\end{figure}

In order to elucidate the possible interplay between inflation and the anomalies matrix rank, we show in Fig.~\ref{infl_rmse} the average number of eigenvalues explaining a portion of the full matrix variance (blue) as well as the analysis RMSE (red), both as a function of the (intentionally fixed, {\it i.e.} non adaptive) inflation factor, for perfect (left panel) and model error (right panel) case with $\sigma_b=0.005$.

Figure~\ref{infl_rmse} suggests that, contrary to our expectations, multiplicative inflation is effective in increasing the rank of the EnKF error covariance matrix. This does not imply, nor it means, that the filter performance improves too. In fact, the RMSE in the perfect model case grows monotonically when the inflation factor exceeds 1.01. Note from Table~\ref{tab:base_exp} that the time-averaged adaptive inflation factor is in this case about 1.01. Similarly, in the model error scenario, the RMSE grows after its minimum at 1.21, which is close to the time-averaged adaptive inflation ({\it cf} Table~\ref{tab:base_exp}).

Multiplicative inflation consists in multiplying the ensemble-based covariance matrix by a scalar coefficient. Thus, how can multiplicative inflation change the covariance matrix's rank? Results in Fig.~\ref{infl_rmse} shed light on an important dynamical property of multiplicative inflation. One would expect at first sight that  multiplicative inflation, as opposed to additive inflation, preserves the span and rank of the anomalies and only affects their amplitude. Nevertheless, even tough the  application of the multiplicative inflation at each individual analysis time does in fact left unchanged the rank and span (it only amounts to as a scalar coefficient multiplying the matrix), its recursive application produces a long-term effect of different nature, altering the span of the covariance matrix. We conjecture this to be a direct consequence of the presence of residual errors in the stable and weakly stable directions. These residual errors would eventually be dampened by model dynamics but in contrast they are amplified on every assimilation cycle by the application of inflation, and then upwelled to the filtered modes, according to the mechanism described in \citet{grudzien18b}. Such a behavior should be more pronounced in the presence of model error, not necessarily due to the increased bursts of instability on the more stable directions, but also due to the required increase of multiplicative inflation. This is confirmed in Fig.~\ref{infl_rmse} by observing that the minimum number of eigenvalues per each variance threshold and the required inflation factor are both larger in the case with model error than in the perfect model case ({\it cf} the blue curves in the two panels). It should be noted that the eigenvalue threshold is approximately the same for each inflation factor setting regardless of the presence of model errors. Therefore multiplicative inflation in sequential ensemble-based DA produces two effects: (i) it  alleviates sampling error sub-estimation of the covariance and represents crudely model error in the filtered subspace and, (ii) it contributes to maintain the filtered subspace dimension to a high value. As shown in Fig.~\ref{infl_rmse}, the rank is as high as $27$ in the model error experiment, in correspondence to an optimal inflation of $\alpha\approx1.21$.

\begin{figure}[!hbt]
	\begin{center}
	\includegraphics[width=0.5\columnwidth]{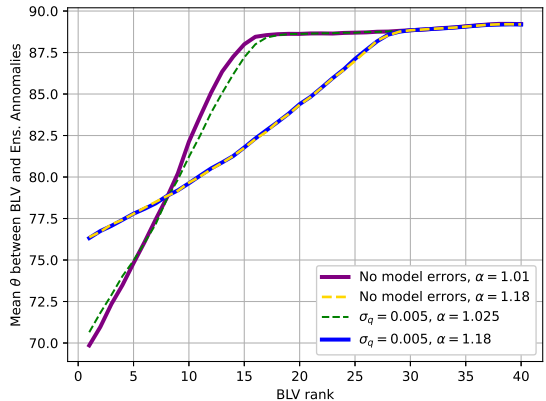}
	\caption{Time averaged angle $\theta$ (degrees) of the ensemble anomalies projected onto the BLVs for the perfect and imperfect ($\sigma_q=0.005$) model scenarios and different fixed multiplicative inflation values ($\alpha$). Results are averaged over $15,000$ assimilation cycles.}
	\label{inflation_comp}
	\end{center}
	\end{figure}

An additional evidence on how multiplicative inflation may be driving the changes in ensemble projections is provided in Fig.~\ref{inflation_comp} by exploring whether the structure of ensemble anomalies depends directly on the changes induced by multiplicative inflation. Figure~\ref{inflation_comp} shows the anomalies projected onto the BLVs for the perfect and imperfect ($\sigma_q=0.005$) model scenarios. In each case, two types of experiments are performed with different multiplicative inflation factors. In the first experiments, the inflation is fixed to the value obtained by averaging the corresponding adaptive inflation values ({\it cf}, Table~\ref{tab:base_exp}). They are $\alpha=1.01$ and $\alpha=1.18$ for the perfect and imperfect model case respectively. In the second experiments, the inflation is set to non optimal values, $\alpha=1.18$ and $\alpha=1.025$ (the latter is the minimum inflation value required to avoid filter divergence) respectively. Results indicate that changes in the alignment of ensemble anomalies onto BLVs are directly related to the amplitude of multiplicative inflation and not necessarily to the presence of model errors. We recall however that the experiments with non optimal inflation are done for the sake of studying the effect on the anomalies alignment onto the BLVs, despite the associate RMSE is larger than in the optimal case. 

Results shown in Fig.~\ref{inflation_comp} confirm our conjecture on the role of multiplicative inflation: it causes changes in the span of the ensemble anomalies. By amplifying error representation outside the unstable-neutral subspace, the multiplicative inflation induces an increase of the ensemble span. This ``artificial'' rank increase does not necessarily imply reductions in the filter state estimation error (the RMSE is much larger than when the optimal or adaptive inflation is used in either case), but it nevertheless reveals of an unknown effect of multiplicative inflation.

\subsection{Reduced-rank model error within the unstable subspace}

In all of the previous experiments, model error was assumed to be full-rank, {\it i.e.} the model error realizations are sampled from a normal distribution with a full-rank covariance matrix $\mathbf{Q}$ given in Eq.~\eqref{eq:Q}. Model error is thus injected on all of the model variables, enhancing the upwelling of error from unfiltered to filtered directions as described in \citet{grudzien18b}. We have also seen that multiplicative inflation prevents the full collapse of the ensemble to the neutral-unstable subspace.

In this section we conduct another set of experiments inspecting the correspondence between the model error rank and EnKF $\mathbf{P}^{\rm a}$, in particular to corroborate whether the rank of $\mathbf{P}^{\rm a}$ would have also changed if model error only spans the unstable-neutral subspace. We devised a proof-of-concept experiment in which all of the model error instantaneous realizations are confined by construction within the subspace spanned by the BLVs associated with non-negative local Lyapunov exponents (LLEs). To achieve this, model error is sampled from $\mathcal{N}({\bf 0},\sigma_q \mathbf{L_k} \mathbf{Q} \mathbf{L}_k)$, where $\mathbf{L}_k=\mathbf{E}^\mathcal{U} {\mathbf{E}^\mathcal{U}}^{\rm T}$ is the orthonormal projector constructed using the local unstable-neutral BLVs, the columns of $\mathbf{E}^i$, where $i=1,...,n_{0k}$ is used to identify that the number of non-negative LLEs fluctuates in time. After being sampled, the norm of model error realizations is rescaled to be consistent with that of the full rank $\mathbf{Q}$ experiments.

\begin{figure}[!ht]
	\begin{center}
    \includegraphics[width=\columnwidth]{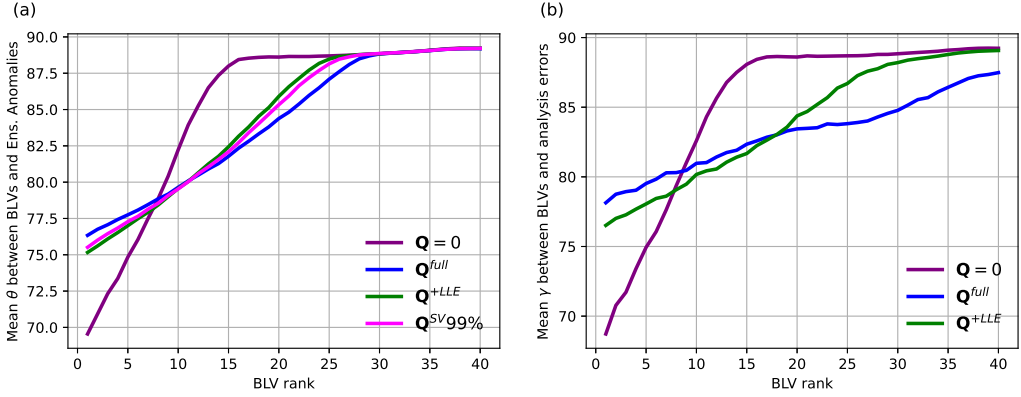}
	\caption{(a) Time averaged angle $\theta$ (degrees) between ensemble anomalies and the BLVs for different configurations of model error. (b) Time and ensemble averaged angle $\gamma$ (degrees) of the analysis errors onto the backward Lyapunov vectors (BLVs) for different representations of model error. Namely, no model error (purple), full rank model error covariance (blue), and model error projected onto locally unstable BLVs (green), model error projected onto decomposed anomalies up to the singular value $s_0$ (pink).}

	\label{err_angle}
	\end{center}
	\end{figure}

By confining the model error within the locally unstable-neutral subspace, the sample forecast covariance is better approximated by the EnKF. This is reflected by a time average RMSE of $0.013$ for the experiment with $\sigma_q=0.005$ compared to $0.016$ obtained in the full-rank case ({\it cf} Table~\ref{tab:base_exp}): RMSE is even slightly below the case of smaller (but full rank) model error, {\it i.e.} $\sigma_q=0.0025$. The time-averaged adaptive multiplicative inflation coefficient is also reduced from $1.18$ to $1.13$. The projection of the ensemble anomalies onto the BLVs (Fig.~\ref{err_angle}a) shows a great similarity with the former full-rank model error experiments (denoted by $\mathbf{Q}^{{\rm full}} $). The current confined model error case is also shown in Fig.~\ref{err_angle}a (denoted by $\mathbf{Q}^{{\rm +LLE}}$). Multiplicative inflation  therefore increases the subspace spanned by the ensemble to new directions which are not directly spanned by model errors. Nevertheless a notable distinction is that the anomalies in the full-rank case  have non-negligible projections on a few more BLVs.

A further understanding of the filter dynamical properties is provided by looking at the mean angle between the analysis error and the BLVs
defined as,
    \begin{equation}
        \gamma^j_k = \arccos \Bigg\{ \frac{(\mathbf{E}^j_k)^\top\boldsymbol{\varepsilon}^{a}_k}{\lVert\boldsymbol{\varepsilon}^{a}_k\rVert}\Bigg\}.
    \end{equation}

 In the perfect model case, the angle between the BLVs and the analysis error (Fig.~\ref{err_angle}b) and that with the ensemble anomalies (Fig.~\ref{err_angle}a) are almost identical, indicating a proper functioning of the EnKF, with the ensemble anomalies properly spanning the directions where the actual error is confined. 
When model error is incorporated, error projections span a broader subspace than anomalies. In the case of model error projecting only on unstable directions, the increased span of analysis errors could be associated to an increased degree of non-linearity due to the presence of errors. In the case with full-rank model errors,  the projections onto the BLVs of the anomalies differ significantly from those of the actual error, particularly in the more stable directions. This indicates in both model error scenarios that the subspace spanned by the ensemble anomalies is not sufficient to capture existing errors, in spite of their source. Error projections onto BLVs of rank 20 up to 25 are particularly noteworthy and it is clear that multiplicative inflation is only able to handle model errors on the leading and weakly stable directions. A different inflation strategy is still needed to treat errors on the more stable portion of the ensemble.

  \section{Hybrid inflation schemes}

  Separating the portion of ensemble anomalies in the span of the BLVs up to the weakly-stable from the vanishing stable directions may enable the design of more sophisticated inflation schemes, targeting additive perturbations to directions that are not spanned by the ensemble. We explore hybrid multiplicative-additive inflation schemes whereby multiplicative inflation is applied to the unstable portion of the spectrum while additive inflation to the remaining part. How can one decompose the anomalies projections to apply such a hybrid approach? A BLV decomposition along the analysis trajectory offers a natural way to achieve this subspace separation. However, explicit computation of BLVs by traditional methods such as QR-decomposition is too expensive and becomes unfeasible for large scale dynamics such as geophysical systems. Given the alignment of the long-term ensemble anomalies onto the unstable-neutral and weakly-stable subspaces described in the previous section, a real time approximation to split the subspaces affected by the hybrid inflation can be based on the anomalies themselves. 
  
  We first decompose the anomalies matrix $\mathbf{A}$ using singular value decomposition (SVD). Under the usual assumption that $N\leq n$, the anomalies matrix $\mathbf{A}$ can be broken down using the reduced singular value decomposition: $\mathbf{A=USV}^\top$, with $\mathbf{U} \in \mathbf{R}^{ n \times N}$, $\mathbf{S} \in \mathbf{R}^{ N \times N}$ and $\mathbf{V}^\top \in \mathbf{R}^{ N \times N}$. The diagonal matrix $\mathbf{S}$ contains the $N$ singular values of $\mathbf{A}$. As shown in Section~\ref{sec_baseline}, $\mathbf P^{\rm a}$ is highly aligned with the neutral-unstable subspace while the remaining variability is associated to (asymptotically) vanishing weakly-stable directions. An identical alignment was found for the filter $\mathbf P^{\rm f}$ matrix. We define $\mathbf{A}^\mathcal{U} \in \mathbf{R}^{n \times N}$ as the matrix containing the projections of the anomalies onto the span of the most unstable portion of the BLV spectrum. This matrix will be referred to as \emph{unstable anomalies} matrix. It is obtained by considering the first $s_0$ columns (rows) of $\mathbf{U}$ ($\mathbf{V}^\top$) in the reduced singular value decomposition. This is, dropping time dependencies,

    \begin{equation}
        \mathbf{A}^\mathcal{U}=\mathbf{U}^{(:,1:s_0)} \mathbf{S}^{(1:s_0,1:s_0)} {\mathbf{V}^{(:,1:s_0)}}^\top,\label{aunst_decomp}
    \end{equation}
    where the threshold $s_0$ is defined as the number of singular values that account for a given threshold of its cumulative sum. The threshold is set to $95\%$. We conducted sensitivity experiments to find that this value is a relatively accurate indicator of the number of BLVs onto which the ensemble anomalies have non-negligible projections over time when using adaptive multiplicative inflation. Indeed when accounting for $95\%$ of cumulative variance, the number of singular values $s_0$ is on average 22 and 13 for the experiments with and without model errors respectively. These numbers are in agreement with the number of leading BLVs onto which ensemble anomalies have non-negligible projections (cf. Section~\ref{sec_baseline}).

    Using Eq.~\ref{aunst_decomp}, we performed an additional idealized experiment in which model error realizations are sampled from $\mathcal{N}(0,\sigma_q \boldsymbol{\Pi_k}^\mathcal{U} \mathbf{Q} {\boldsymbol{\Pi_k}^\mathcal{U}})$, where $\boldsymbol{\Pi_k}^\mathcal{U}=\mathbf{A}_k^\mathcal{U} {\mathbf{A}_k^\mathcal{U}}^{+}$ is a unitary operator and is the orthogonal projector onto the column space of ${\mathbf{A}_k^\mathcal{U}}^{+}$, with  ${\mathbf{A}_k^\mathcal{U}}^{+}$ being the Moore-Penrose pseudo-inverse of ${\mathbf{A}_k^\mathcal{U}}$. In this case, model error instances are ensured to belong to the span of the ensemble. Not surprisingly, the RMSE is almost identical to the assimilation experiments in which model error instances were projected onto the span of the locally unstable BLVs. The mean angle between ensemble anomalies and BLVs is shown in Fig.~\ref{err_angle}a (light purple line).
    
    The remaining $(N-s_0)$ columns (rows) of $\mathbf{U}$ ($\mathbf{V}^\top$) are used to define the \emph{stable anomalies} matrix:

    \begin{equation}
        \mathbf{A}^\mathcal{S}=\mathbf{U}^{(:,s_0+1:N)} \mathbf{S}^{(s_0+1:N,s_0+1:N)} {\mathbf{V}^{(:,s_0+1:N)}}^\top.
    \end{equation}

    By definition, matrices $\mathbf{A}^\mathcal{U}$ and $\mathbf{A}^\mathcal{S}$ are mutually orthogonal ($n \times N$) matrices of rank $s_0$ and $(N-s_0)$ respectively, so the anomalies matrix at time $t_k$ can be written as
    \begin{equation}
    \mathbf{A}_k=\mathbf{A}_k^\mathcal{U}+\mathbf{A}_k^\mathcal{S},
    \end{equation}

    and,

    \begin{equation}
    (N-1)\mathbf{P}^f_k=\mathbf{A}_k^{\mathcal{U}}{\mathbf{A}_k^\mathcal{U}}^\top+\mathbf{A}_k^\mathcal{S}{\mathbf{A}_k^\mathcal{S}}^\top.
    \label{eq:pf}
    \end{equation}

\begin{figure}[!hbt]
	\begin{center}
    \includegraphics[width=\columnwidth]{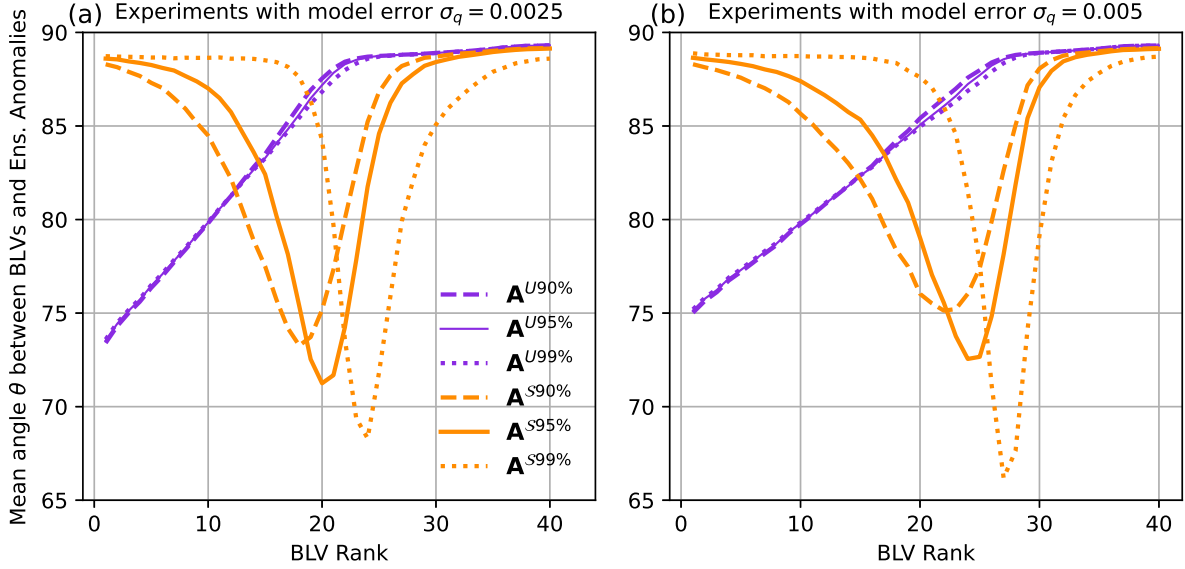}
	\caption{Time and ensemble averaged angle $\theta$ (degrees) of the dynamically unstable (purple) and stable (orange) portion of ensemble anomalies $\mathbf{A}^U$ (purple) onto the backward Lyapunov vectors for experiments with model errors (a) $\sigma_q=0.0025$ and (b) $\sigma_q=0.005$. Multiple thresholds for the amount of total variance in  $\mathbf{P}^{\rm f}$  accounted by unstable anomalies are shown (i.e. ${90\%,95\%,99\%})$.}
	\label{blv_au_as}
	\end{center}
	\end{figure}

    Stable and unstable anomalies projections onto BLVs are shown in Fig.~\ref{blv_au_as} using the experimental settings defined in Section~3 for the same model error configurations ({\it i.e.}, $\sigma_q=0.0025$ and $\sigma_q=0.005$). In each case $s_0$ is dynamically adjusted to account for the corresponding percentage of $\mathbf{P}^{\rm f}$ total variance and the multiplicative inflation is adaptively adjusted. Anomalies associated to $A^\mathcal{U}$ are not only constrained to BLVs associated to asymptotically unstable LLEs, but indeed they span a subspace of higher dimensions. This subspace is increased or decreased based on the level of multiplicative inflation associated to each configuration of model error (see Table~\ref{tab:base_exp} for reference).  In contrast, $A^\mathcal{S}$ anomalies are largely aligned with higher order BLVs. In fact the larger $s_0$ the higher the order of the BLVs where the residual variance is explained (see how the corresponding yellow lines shift toward right in both panels of Fig.~\ref{blv_au_as}). It should be noticed that  $A^\mathcal{S}$ accounts for a tiny portion of the total variance of $\mathbf{P}^{\rm f}$, representing residual ensemble anomalies that lay mostly outside of the unstable subpsace.
    
    Using the proposed decomposition, it is possible to replicate the span of the locally unstable and weakly stable BLVs, to which one has not usually access in realistic high-dimensional cases due to their prohibitive computational cost. 
    However, this type of decomposition depends on the chosen total variance threshold, which impacts on $s_0$. As model error increases, the optimal multiplicative inflation needed will also increase. Therefore, it is expected that the proportion of anomalies projecting onto stable BLVs will increase. Consequently, larger values of $s_0$ are required to properly disentangle anomalies. Moreover, the possible incorporation of additive inflation through random perturbations may have an instantaneous effect similar to the introduction of model errors and will also lead to a shift toward higher values of $s_0$ for a given variance threshold. These type of feedback processes will drive the need for further threshold adjustments which are outside of the scope of this work. 
    
    The decomposition in Eq.~\ref{eq:pf} allows us to design inflation schemes that operate distinctively over the portions of the covariance matrix aligned onto the unstable and stable BLVs. We choose to inflate the unstable anomalies with a multiplicative inflation scheme such that $\mathbf{A}_k^{'\mathcal{U}}=\sqrt{\rho}\mathbf{A}_k^{\mathcal{U}}$.
    Conversely, for the stable portion of the spectrum, we propose an additive inflation mechanism that attempts to account for errors accumulating on this portion of the spectrum as shown in Fig.~\ref{err_angle}b (blue line). Additive inflation is thus incorporated on directions that are orthogonal to the leading $s_0$ anomalies directions but not necessarily limited to the remaining $N-s_0$ anomalies directions, following  model error covariance $\mathbf{Q}$. The most unstable directions are properly inflated by the effect of multiplicative inflation alone. We propose two hybrid additive-multiplicative inflation schemes that are the content of the following sections.

    \subsection{Hybrid projected stochastic inflation}

    Additive covariance inflation schemes ({\it e.g.} \citealp{whitaker08}, \citealp{hamill05}) consists of drawing random perturbations from a specified error distribution, and adding them to the (prior) forecast ensemble before being updated. Usually these perturbations do not take into account model dynamics and are usually sampled from a full rank covariance matrix. In contrast, we are interested in sampling them from a subspace orthogonal to the unstable BLVs. Using the non-stable anomalies projector defined in the previous section, ${\boldsymbol{\Pi}_k^\mathcal{U}=\mathbf{A}^\mathcal{U} {\mathbf{A}^\mathcal{U}}^{+}}$, random samples can be drawn that are orthogonal to the non-stable modes. The inflated anomalies used for ETKF analysis calculations are then
    
    \begin{equation}
    \mathbf{A_k}'=\sqrt{\rho} \mathbf{A}_k^\mathcal{U}+\mathbf{A}_k^\mathcal{S}+\sqrt{\sigma_q}(\mathbf{I} - \boldsymbol{\Pi}_k^\mathcal{U}) \mathbf{Q}^{1/2} \boldsymbol{\Xi}_k,
    \label{addensinfl}
    \end{equation}
    with $\boldsymbol{\Xi}=\{\boldsymbol{\xi}_1,\boldsymbol{\xi}_2,\ldots,\boldsymbol{\xi}_N\}$, and all $\boldsymbol{\xi}_i$ are centered independent random draws from $\mathcal{N}(0,\mathbf{I}_n)$. By construction, the last two terms in Eq.~\ref{addensinfl} span a subspace orthogonal to the unstable and weakly-stable anomalies subspace spanned by the ensemble. The introduction of stochastic noise not aligned with the unstable anomalies leads to a slightly more uniform eigenspectrum for the covariance matrix. These changes have a direct impact on the value of $s_0$, which is now increased proportionally with the amplitude of stochastic additive inflation. To account for this, we set $s_0$ as the number of singular values required to achieve at least $90\%$ of cumulative percentage sum of singular values. 

    Perturbations associated to non-stable modes are inflated through a multiplicative inflation scheme instead (first term on the r.h.s. of Eq.~\ref{addensinfl}). The adaptive multiplicative inflation scheme of \citet{miyoshi11} was adapted to this framework in order to scale the inflation coefficient only by taking into account the non-stable portions of the spectrum. Details of this modified adaptive scheme are given in Appendix A. 

    \subsection{Hybrid projected deterministic inflation}

    Our second hybrid approach is inspired by the work of \citet{raanes15}. In this approach the additive inflation acts on the forecast step to account model errors via a deterministic transformation of the ensemble anomalies.  By doing so, it removes the randomness  associated to the stochastic sampling step of the additive  inflation \cite{hamill05}. This approach is specifically designed for square-root EnKFs \citep{tippett03}. The base implementation, named SQRT-CORE in \citet{raanes15}, aims to inflate the covariance matrix consistently with the model error covariances according to

    \begin{equation}
        \mathbf{P}^{\rm f} = \frac{1}{(N-1)}\mathbf{A}\mathbf{A}^\top+\mathbf{A}\mathbf{A}^{+} \mathbf{Q}\mathbf{A}\mathbf{A}^{+},
        \label{eq:pf_det}
      \end{equation}
    where temporal dependencies on Eq. (\ref{eq:pf_det}) were dropped for conciseness. The propoposed formulation ensures that covariance is inflated only in the span of the ensemble anomalies, however only a portion of model error $\mathbf{Q}$ is accounted with this formulation. To satisfy Eq.~(\ref{eq:pf_det}), \citet{raanes15} proposed the following (symmetric) square root formulation for the anomalies,

    \begin{equation}
        \mathbf{A}'=\mathbf{AT} = \mathbf{A}[\mathbf{I}_N+(N-1)\mathbf{A}^{+} \mathbf{Q}(\mathbf{A}^{+})^\top]^{1/2}.
    \end{equation}

     Based on the above, we propose an hybrid inflation scheme similar in spirit to the one  in Section~4.1 but that avoids the stochastic term. This second approach, named {\it hybrid projected deterministic inflation}, consists in confining the deterministic inflation scheme SQRT-CORE from \citet{raanes15} to the span of the stable anomalies. This is,

    \begin{equation}
        \mathbf{A'}^\mathcal{S}= \mathbf{A^\mathcal{S}}[ \mathbf{I}_N+\sigma_q (N-1)\mathbf{A^\mathcal{S}}^{+} \mathbf{Q}(\mathbf{A^\mathcal{S}}^{+})^\top]^{1/2}.
    \end{equation}

    The inflated covariances can be reconstructed as in Eq. (\ref{eq:pf}), since again inflated anomalies $\mathbf{A'}^\mathcal{S}$ belong to the span of $\mathbf{A}^\mathcal{S}$ and are thus orthogonal to $\mathbf{A}^\mathcal{U}$. The threshold $s_0$ is defined using the same criterion as in the projected stochastic inflation scheme of the previous section. For anomalies aligned with non-stable modes an adaptive multiplicative covariance inflation scheme is used. The resulting inflated anomalies are then

    \begin{equation}
    \mathbf{A_k}'=\sqrt{\rho} \mathbf{A}_k^\mathcal{U}+\mathbf{A'}_k^\mathcal{S}.
    \label{detensinfl}
    \end{equation}

    We emphasize that the hybrid projected deterministic inflation in Eqs.~(14)--(15) differs from the hybrid projected stochastic inflation of Section~4.1 not only due to its deterministic nature, but most importantly due to additive inflation being constrained to the subspace of stable anomalies spanned by the ensemble. It also distinguishes from standard multiplicative inflation by incorporating perturbations that are proportional to $\mathbf{Q}$ and are completely aligned with the most stable directions of the ensemble.

    \subsection{Comparing the hybrid inflation schemes}

    The proposed hybrid inflation schemes are compared numerically under the model configuration of Section~3 with $\sigma_q=0.005$. The hybrid schemes are compared against two baselines: the adaptive multiplicative inflation scheme from \citet{miyoshi11} and the additive inflation scheme in which forecasts are inflated additively with random perturbations sampled from $\mathcal{N}(0,\mathbf{Q})$. For a more fair comparison, random perturbations from the additive inflation scheme are scaled by a factor of $\gamma=1.18$, this value was obtained from an exhaustive search to optimize RMSE. 
    
    In hybrid schemes, the multiplicative inflation applied to the unstable portion of anomalies is estimated adaptively at each assimilation cycle using the approach in \citet{miyoshi11}. We also include the results using the {\it SQRT-DEP} scheme. An extension of the additive inflation SQRT- CORE scheme proposed by \citet{raanes15} that accounts for stochastic perturbations sampled from outside the ensemble subspace plus a stochastic correction for potential residual effects within the ensemble subspace (see \citealp{raanes15}, their Section~6). The RMSE, the additive inflation scaling factor, and the mean optimal multiplicative inflation factor for each scheme are given in Table~\ref{tab:hybrmse}. The RMSE values achieved using the different schemes are very similar, although the hybrid projection deterministic scheme performs slightly better than its stochastic counterpart. This suggests that mitigating sampling errors in the additive inflation term is beneficial. Along this line, we see that compared to SQRT-DEP, the hybrid stochastic schemes are not able to achieve the same level of RMSE, emphasizing the impact of sampling errors associated to additive inflation schemes.
     
\begin{figure}[!hbt]
	\begin{center}
    \includegraphics[width=\columnwidth]{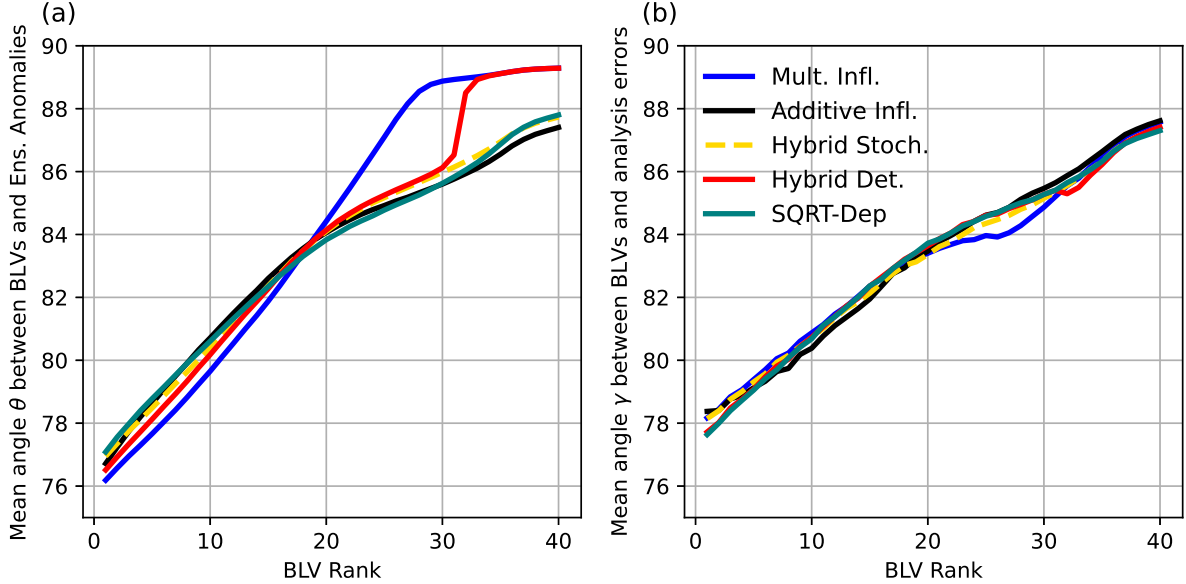}
	\caption{(a) Time and ensemble averaged angle $\theta$ (degrees) of the ensemble anomalies projected onto the backward Lyapunov vectors (BLVs) for the multiplicative inflation scheme (blue), standard additive inflation scheme (black),  hybrid projected stochastic inflation (yellow), hybrid projected deterministic inflation (red) and SQRT-DEP (green). (b) Time and ensemble averaged angle $\gamma$ (degrees) of the analysis errors projected onto the backward Lyapunov vectors (BLVs) for the same assimilation configurations}
	\label{blv_hyb}
	\end{center}
	\end{figure}

The hybrid schemes have an average value of $s_0=22$ and $s_0=24$ for the stochastic and deterministic cases respectively. This implies that a larger portion of the  anomalies space is treated via multiplicative inflation in the deterministic version. The projection of ensemble anomalies onto BLVs is shown in Fig.~\ref{blv_hyb}a. The hybrid schemes show a very similar projection onto BLVs of rank $r<22$, indicating that the differences in the value of $s_0$ for the hybrid schemes may not play a dominant role. For the multiplicative scheme, the projection onto higher order BLVs is progressively reduced, and the reduction ratio remains quite uniform. In contrast, the introduction of inflation in directions which are orthogonal to the unstable BLVs and helps to retain more effectively perturbations that align with BLVs of rank $r>20$ onto the subspace spanned by unstable and weakly stable BLVs. The hybrid projected deterministic schemes incorporates inflation only in the subspace spanned by the ensemble, hence the anomalies covariance matrix cannot span the full subspace spanned by the BLVs (see red line in Fig.~\ref{blv_hyb}a). On the other hand,  inflation in the hybrid stochastic inflation scheme is projected onto the part of the ensemble associated with the non-unstable modes, while consistently sampling from $\mathbf{Q}$. The alignment of anomalies with  BLVs for this scheme holds a large resemblance with the SQRT-DEP method. This is related to the fact that both schemes not only inflate ensemble anomalies, but they also combine it with random noise sampled from a subspace orthogonal to the ensemble subspace. As a comparison, the additive inflation scheme leads to covariances with about the same degree of projection onto BLVs. However, this type of projections analysis ignores the contribution of anomalies which are orthogonal to BLVs subspace, which may be responsible for its poorer performance in terms of RMSE. 
    
    Analysis errors projected onto the BLVs do not differ significantly among the hybrids and SQRT-DEP schemes. All schemes that explicitly inflate directions outside the leading unstable and weakly BLVs reduce errors by projecting on higher order BLVs when compared to multiplicative inflation alone.

	\begin{table}[!hbt]
		\begin{center}
		\caption{RMSE for several inflation schemes}
		\label{tab:hybrmse}
        \begin{tabular}{|l|c|c|c|c}
			\hline
			 & RMSE & Unstable    & Stable   \\
 			 &    &     inflation  factor &  inflation factor \\
			\hline
			Multiplicative inflation & 0.0163 & $\alpha=$1.18 & $\alpha=$1.18  \\
			\hline
			Additive inflation & 0.0177 & $\gamma=$1.18 & $\gamma=$1.18  \\
			\hline
			Hybrid proj. stochastic inflation & 0.0160 & $\alpha=$1.13 & $\alpha=$1.00 \\
			\hline
		    Hybrid proj. deterministic inflation & 0.0152 & $\alpha=$1.11 & $\alpha=$1.00 \\
			\hline
			SQRT-DEP & 0.0151 & $\alpha=$1.00 & $\alpha=$1.00 \\
			\hline

        \end{tabular}
		\end{center}
	\end{table}

\begin{figure}[!hbt]
	\begin{center}
	\includegraphics[width=\columnwidth]{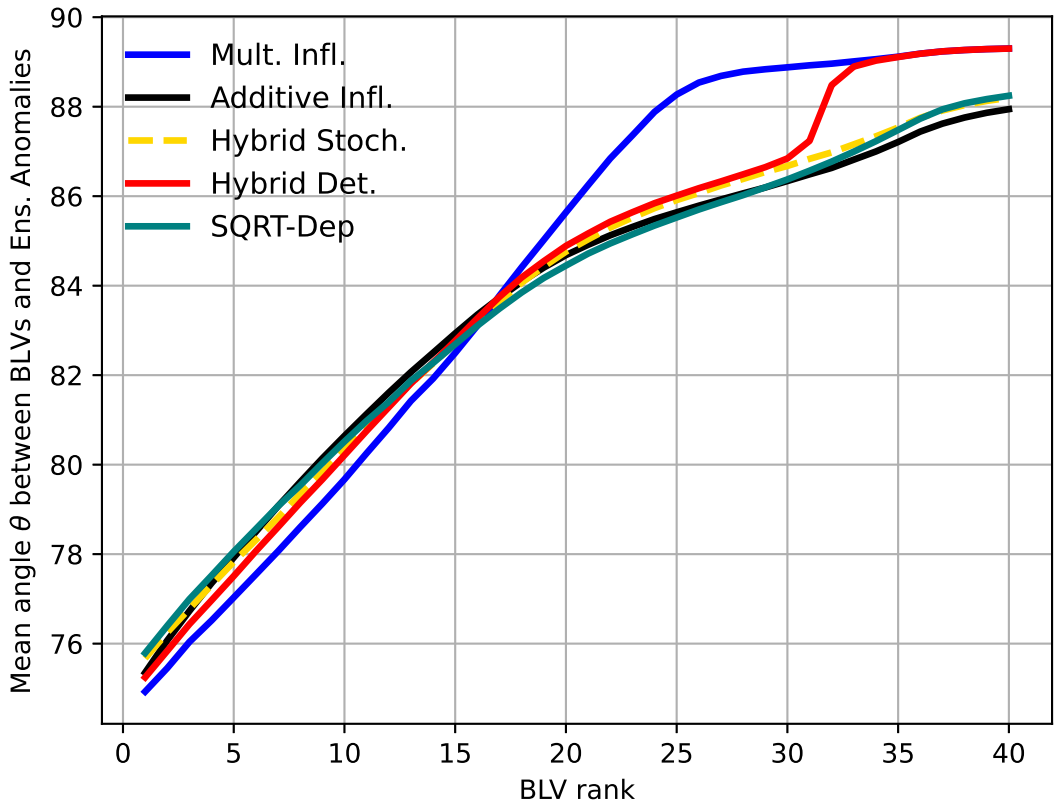}
	\caption{Same as in \ref{blv_hyb} but assuming observations are only available at every other grid point.}
	\label{blv_hyb_H2}
	\end{center}
	\end{figure}

 The proposed experimental framework is intentionally characterized by a strong observational constraint. We performed an additional set of experiments with half of the observations, namely we observe only every other variable. The reduced density of observations alleviates the need of multiplicative inflation across all of the experiments with adaptive inflation ({\it cf.} Table~\ref{tab:hybrmse_sparse}). The analysis RMSE improvement using the hybrid schemes was around $10\%$ with respect to multiplicative inflation, which is very close to the results obtained in the fully observed case. While the ensemble anomalies projection shown in Fig.~\ref{blv_hyb_H2} retains several features from the previous experiment, it should be noticed that the angle of projection onto the stable modes is much smaller in this case. For the case with only multiplicative inflation, the ensemble anomalies span a reduced subspace of the BLV spectrum. This may indicate that the ensemble is constrained to a reduced subspace where model dynamics play a more dominant role than model errors.

	\begin{table}[!hbt]
		\begin{center}
		\caption{RMSE for several inflation schemes - Reduced observations network}
		\label{tab:hybrmse_sparse}
        \begin{tabular}{|l|c|c|r|r}
			\hline
			 & RMSE & Unstable    & Stable   \\
 			 &    &     inflation  factor &  inflation factor \\
			\hline
			Multiplicative inflation & 0.0221 & $\alpha=$1.14 & $\alpha=$1.14  \\
			\hline
			Hybrid proj. stochastic inflation & 0.0217 & $\alpha=$1.09 & $\alpha=$1.00 \\
			\hline
		    Hybrid proj. deterministic inflation & 0.0202 & $\alpha=$1.07 & $\alpha=$1.00 \\
			\hline
			SQRT-DEP & 0.0201 & $\alpha=$1.00 & $\alpha=$1.00 \\
			\hline
		\end{tabular}	
        
		\end{center}
	\end{table}

\section{Discussions}

    In this work, the effects multiplicative and additive covariance inflation over ensemble dynamics are examined within ensemble-based data assimilation. To better understand the role of inflation, we  evaluated its impact on the span of ensemble anomalies. A previous work \citep{bocquet2017a} proved that filter stability is ensured only when the long-term unstable subspace spanned by the BLVs is represented by the filter. Following \citet{bocquet2017a} methodology, we found that  the introduction of model error leads to changes in the alignment of ensemble anomalies onto the BLVs for a highly observed system. An evaluation of the geometric angle for this alignment shows that the ensemble of anomalies is also projected onto weakly-stable BLVs. 

    One of the most important results obtained in this work is the fact that the use of multiplicative inflation alone is enough to retain anomalies in the span of weakly stable BLVs, regardlessly of the presence of model error. The number of weakly-stable BLVs that support the Kalman filter analysis covariance matrix seems to be dynamically adjustable via multiplicative inflation. This mechanism may explain the success of multiplicative inflation schemes in accounting for model errors that typically lay outside of the unstable subspace that supports the ensemble, in line with the findings in \citet{grudzien18b} for the Extended Kalman Filter. The exact mechanisms that drive the increase of the subspace spanned by the anomalies is not yet clear. We hypothesize that it could be a result of residual anomalies on unfiltered directions that cannot be effectively dampened by the filter due to the effect of multiplicative inflation.

    Assuming that the unstable and leading weakly-stable modes can be represented by means of a properly tuned multiplicative inflation scheme, we propose two methods that attempt to represent and incorporate model errors associated to the remaining asymptotically stable modes. In this way, we divide the treatment of sampling errors on the unstable modes from the need for representing the model errors effects on the stable subspace. The proposed methods are based on a singular value decomposition of the forecast anomalies to isolate the components associated to the more stable modes. The proposed methods aid to understand the impact of additive inflation on the presence of model error. An alternative approach to exploit the subspace orthogonal to the ensemble space through a variational correction was proposed by \citet{chang20} in the scope of hybrid-gain data assimilation.

    Our results suggest that errors are expected to have a more uniform projection onto the BLVs as model error increases. It may be interesting to evaluate the performance of the hybrid schemes with an increased level of non-linearity. In that scenario, the representation of errors outside of the unstable modes may still have a critical role. It is possible, nevertheless, that multiplicative inflation alone is enough to align the filter towards the subspace spanned by a suitable number of BLVs in that case too.

\section{Appendix A - Adaptive multiplicative inflation for unstable subspace}

The adaptive multiplicative covariance inflation scheme used in this work is outlined in \citet{miyoshi11}. This scheme exploits one of the diagnostic relationships described by \citet{desroziers05}, namely

\begin{equation}
        {\langle \mathbf{d}\mathbf{d}^{\rm T}\rangle}=\mathbf{HB}\mathbf{H}^{\rm T}+\mathbf{R},
\end{equation}

where $\mathbf{d}=\mathbf{y}-\mathbf{Hx}^{\rm b}$ are the observation minus background innovations and $\mathbf{B}$ is the background covariance matrix. This relationship is valid as long as the observation and background errors are perfectly specified. For the ensemble Kalman filter, the equation can be approximated by incorporating an inflation factor to the ensemble derived forecast errors covariance matrix. 

\begin{equation}
        {\langle \mathbf{d}\mathbf{d}^{\rm T} \rangle} = \alpha  \mathbf{HP}^{\rm b}\mathbf{H}^{\rm T}+\mathbf{R},
\end{equation}

Based on diagonals terms, the optimal instantaneous coefficient can be inferred as

\begin{equation}
       \alpha = \frac{{\langle \mathbf{d}\mathbf{d}^{\rm T}\rangle} - \rm tr(\mathbf{R})}{\rm tr(\mathbf{HP}^{\rm b}\mathbf{H}^{\rm T})},
      \label{adapt_infl_eq} 
\end{equation}

To adjust the adaptive multiplicative inflation factor to the unstable subspace we modify Eq. \ref{adapt_infl_eq} in order to bound it to the unstable subspace

\begin{equation}
       \alpha^\mathcal{U} = \frac{ {\langle \mathbf{d}\mathbf{d}^{\rm T}\rangle}-{\rm tr(\mathbf{HA}^{\mathcal{S}}\mathbf{A}^{{\mathcal{S}}{\rm T}}\mathbf{H}^{\rm T})/(N-1)} - \rm tr(\mathbf{R})}{\rm tr(\mathbf{HA}^{\mathcal{U}}\mathbf{A}^{\mathcal{U}}{\rm T}\mathbf{H}^{\rm T})/(N-1)},
      \label{adapt_infl_eq_unst} 
\end{equation}

Instantaneous estimations of inflation factor should however take into account uncertainties associated to sampling and unsuitable definition of $\mathbf{R}$ -the latter is not an issue in our synthetic setup. For this reason, the inflation factor is updated using a scalar Kalman filter ensuring also a smooth convergence towards an optimal value. A detailed discussion of parameter update process is given by \citet{li09} and also by \citet{miyoshi11}.

\bibliography{main}

\end{document}